\documentclass[useAMS,usenatbib,babel]{mn2e}

\usepackage[english,english]{babel}
\usepackage{amsmath}
\usepackage{amssymb,amsfonts,textcomp}
\usepackage{array}
\usepackage{supertabular}
\usepackage{hhline}
\usepackage{hyperref}
\usepackage[usenames]{color}

\usepackage{textcomp}

\usepackage{graphicx}

\usepackage{dcolumn}
\usepackage{bm}
\usepackage{comment}
\usepackage{color}

\newcommand{\mD}{{\cal D}}
\newcommand{\mP}{{\cal P}}

\newcommand{\vx}{\textbf{x}}

\newcommand{\dd}{\hbox{d}}
\newcommand{\ii}{\hbox{i}}

\newcommand{\Det}{{\rm Det}}

\newcommand{\mgen}{{\cal M}}
\newcommand{\hrho}{{\hat\rho}}

\begin{document}
\title[ Statistics of densities and slopes in count-in-cells]{The joint statistics of mildly non-linear cosmological densities and  slopes in count-in-cells}
\author[F.~ Bernardeau, S.~ Codis and C.~Pichon]{
\parbox[t]{\textwidth}{
Francis Bernardeau$^{1,2}$, Sandrine Codis$^{1}$ and Christophe Pichon$^{1}$}
\vspace*{6pt}\\
\noindent$^{1}$CNRS \& UPMC, UMR 7095, Institut d'Astrophysique de Paris, F-75014, Paris, France\\
$^{2}$ CNRS \& CEA, UMR 3681, Institut de Physique Th\'eorique, F-91191 Gif-sur-Yvette, France
}

\maketitle
\begin{abstract}
In the context of count-in-cells statistics, the joint probability distribution of the  density in two concentric spherical shells is predicted from first first principle for sigmas of the order of one. The agreement with simulation is found to be excellent. This statistics allows us to deduce the conditional one dimensional  probability distribution  function of the  slope  within under dense (resp. overdense) regions, or of the density for positive or negative slopes. The former conditional distribution is likely to be more robust 
in constraining the cosmological parameters as the underlying dynamics is less evolved in such regions.
A fiducial dark energy experiment is implemented on such counts derived from $\Lambda$CDM simulations.
\end{abstract}

\section{Introduction}

With the advent of large galaxy surveys (e.g. SDSS and in the coming years Euclid~\citep{Euclid}, LSST), astronomers have ventured into the era of statistical cosmology and big data. Hence, there is a dire need for them to build tools that can efficiently extract as much information as possible from these huge data sets at high and low redshift. In particular, this means being able to probe the non-linear regime of structure formation.
The most commonly used tools to extract statistical information from the observed galaxy distribution are N-point correlation functions \citep[e.g][]{Scoccimarro98} which quantify how  galaxies are clustered. In our initially Gaussian Universe the matter density field is fully described by its two-point correlation function. However departure from Gaussianity occurs when the growth of structure becomes non-linear (at later times or smaller scales), providing information
that is not captured by the two-point correlation function but is recorded in part in the three-point correlation function. Obviously N-point correlation functions are increasingly difficult to measure when N increases. They are noisy, subject to cosmic variance and highly sensitive to systematics such as the complex geometry of  surveys.  It is thus essential to find alternative estimators to extract information from the non-linear regime of structure formation in order to complement  these classical probes. This is in particular critical if we are to understand the origin of dark energy, which accounts  for $\sim$ 70 \% of the energy budget of our Universe.

One such method to accurately probe the non-linear regime is to implement {perturbation theory} in a highly symmetric configuration (spherical or cylindrical symmetry) for which the full joint cumulant generating functions can be constructed. Such constructions take advantage of the fact that 
non-linear solutions  to the gravitational dynamical equations  (the so-called spherical collapse model) are known exactly. Corresponding observables, such as galaxy counts in concentric spheres or discs,  then yield very accurate analytical predictions in the mildly non-linear regime, well beyond what is usually achievable using other estimators. 
The corresponding symmetry implies that the most likely dynamical evolution (amongst all possible mapping between the initial and final density field) 
is that corresponding to the spherical  collapse for which we can write an explicit linear to nonlinear mapping.
This has been demonstrated in the limit of zero variance using direct diagram resummations~\citep{1992ApJ...390L..61B,1994A&A...291..697B}\footnote{The original derivations were actually derived from 
the hierarchical model that aimed at describing the fully nonlinear regime, \citep{1989A&A...220....1B}} which was later shown to correspond to a saddle approximation~\citep{1993ApJ...412L...9J,2002A&A...382..412V,Bernardeau14}. The key point on which this whole paper is based upon, is that 
the zero variance limit is shown to provide a remarkably good working model for finite variances \citep{1994A&A...291..697B,Bernardeau14}.  

This formalism 	also allows to weigh non-uniformly different regions of the universe making possible to take into account the fact that
the noise structure in surveys is not homogenous. For instance, low density regions are probed by fewer galaxies.
Conversely, on dynamical grounds, we also expect the level of non-linearity in the field to be in-homogenous: low density regions are less non-linear. 
Hence it is of interest to build statistical estimators which probe the mildly non linear regime and that can be tuned to probe 
subsets of the field, offering the best compromise between these constraints. In the context of the cosmic density field, the construction of conditional
distributions naturally leads to the elaboration of joint probability distribution functions (PDF hereafter) of the density in concentric cells.

Following \cite{Bernardeau14} (hereafter BPC), we propose in Section 2 to extend one-point statistics of density profiles and to the full joint probability distribution function of the density in two concentric spheres of different radii. This is obtained using perturbation theory core results on the
cumulant generating function, the double inverse Laplace transform of which is then computed from 
brute force numerical integration. From that PDF, we will also present the statistics of density profiles restricted to underdense (resp. overdense) regions, and the statistics of density  restricted to positive (resp. negative) slopes (Section 3). Theoretical predictions will be shown to be in very good agreement with simulations in the mildly non-linear regime. Dependence with redshift will also be discussed.
Finally Section 4 presents a simple fiducial dark energy experiment, while 5 wraps up.

\section{The 2-cell density statistics}
For the sake of clarity, let us present and briefly comment the formalism. We
 consider two spheres ${\cal S}_{i}$ of radius $R_{i}$ ($i=1,2$) centered on a given location of space $\vx_{0}$. Our goal is to derive the joint PDF of the density in ${\cal S}_{1}$ and ${\cal S}_{2}$ denoted $\hrho_{i}$ and rescaled so that $\left\langle \hrho_{i}\right \rangle=1$. 

\subsection{The cumulant generating function}
In the cases we are interested in, the joint statistical properties of $\hrho_{1}$ and $\hrho_{2}$ are fully encoded in their moment generating function
\begin{eqnarray}
\mgen_{R_{1}R_{2}}(\lambda_{1},\lambda_{2})&=& \sum_{p,q=0}^{\infty}\ \langle \hrho_{1}^{p} \hrho_{2}^{q}\rangle \frac{\lambda_{1}^{p}\lambda_{2}^{q}}{p!\,q!}\,,
\\
&=&\langle \exp(\lambda_{1}\hrho_{1}+\lambda_{2}\hrho_{2})\rangle\label{eq:mgen}
\,,
\end{eqnarray}
that can be related to the {\sl cumulant} generating function, $ \varphi_{R_{1}R_{2}}(\lambda_{1},\lambda_{2})$, through
$\mgen_{R_{1}R_{2}}(\lambda_{1},\lambda_{2})=\exp\left[ \varphi_{R_{1}R_{2}}(\lambda_{1},\lambda_{2})\right]$,
so that
\begin{eqnarray}
\exp\left[ \varphi_{R_{1}R_{2}}(\lambda_{1},\lambda_{2})\right]&&\nonumber\\
&&\hspace{-3cm}=\int\dd\hrho_{1}\dd\hrho_{2}
\mP_{R_{1}R_{2}}(\hrho_{1},\hrho_{2}) \exp(\lambda_{1}\hrho_{1}+\lambda_{2}\hrho_{2}),
\label{Mint}
\end{eqnarray}
where
$\mP_{R_{1}R_{2}}(\hrho_{1},\hrho_{2})$ is the joint PDF of having density $\hrho_{1}$ in ${\cal S}_{1}$ and $\hrho_{2}$ in ${\cal S}_{2}$. 
We will now exploit a theoretical construction that permits the explicit calculation of $\mP_{R_{1}R_{2}}(\hrho_{1},\hrho_{2})$.
\subsubsection{Upshot}
As we will sketch in the following, this theoretical construction yields the explicit time dependence of the \textsl{Legendre transform} of $\varphi_{R_{1}R_{2}}(\lambda_{1},\lambda_{2})$ in the quasi-linear regime. 
Such a Legendre transform is defined as
\begin{equation}
\Psi_{R_{1}R_{2}}(\rho_{1},\rho_{2})=\lambda_{1}\rho_{1}+\lambda_{2}\rho_{2}-\varphi_{R_{1}R_{2}}(\lambda_{1},\lambda_{2}),\label{phifromLeg}
\end{equation}
where $\rho_{i}$ are determined implicitly by the stationary conditions
\begin{equation}
\lambda_{i}=\frac{\partial}{\partial \rho_{i}} \Psi_{R_{1}R_{2}}(\rho_{1},\rho_{2})\,, \quad {i=1,2}
\,.\label{statCond3}
\end{equation}
The fundamental relation is then that, in the limit of zero variance, this Legendre transforms taken at two different times, 
$\Psi(\rho_{1},\rho_{2};\eta)$ and $\Psi'(\rho_{1},\rho_{2};\eta')$,
take the {\sl same} value
\begin{equation}
\Psi_{R_{1}R_{2}}(\rho_{1},\rho_{2};\eta)=\Psi_{R'_{1}R'_{2}}(\rho'_{1},\rho'_{2};\eta')\,,
\label{FondRel}
\end{equation}
provided that
$\rho_{i}R_{i}^{3}=\rho'_{i}{R'_{i}}^{3}
$,
and that 
$\rho'_{i}$ and $\rho_{i}$ are  linked together through  the nonlinear dynamics of spherical collapse.

Equation~(\ref{FondRel}), when applied to an arbitrarily early time $\eta'$, yields a relation between $\Psi(\rho_{1},\rho_{2};\eta)$ and the 
statistical properties of the {\sl initial} density fluctuations. In particular, for Gaussian initial conditions, $\Psi(\rho_{1},\rho_{2};\eta_{i})$ 
can easily be calculated and expressed in terms of elements of covariance matrices,
\begin{equation}
\Psi_{R_{1}R_{2}}(\rho_{1},\rho_{2};\eta_{i})=\frac{1}{2}\sum_{i,j\le2}\Xi_{ij}(\rho_{i}-1)(\rho_{j}-1)\,,
\end{equation}
where $\Xi_{ij}$ is the {\sl inverse} of the matrix of  covariances, $\Sigma_{ij}=\langle\tau_{i}\tau_{j}\rangle$, between the initial density contrasts 
in the two concentric spheres of radii $R_{i}$.
One can then write the cumulant generating  function at any time
through the spherical collapse mapping
between one final density at time $\eta$ in a sphere of radius $R_{i}$ and one initial contrast in a sphere centered on the same point
 and with radius $R'_i=R_{i}\rho_{i}^{1/3}$ (so as to encompass the same total mass);
 it can be written formally as 
\begin{equation}
\rho_{i}=\zeta_{\rm SC}(\eta,\tau_{i})
\approx \frac{1}{(1-D_{+}(\eta)\tau/\nu)^{\nu}}\,,\label{zetarelation}
\end{equation}
where, for the sake of simplicity, we use here a simple prescription,
with $D_{+}(\eta)$ the linear growth factor and $\nu=21/13$ to reproduce the high-z skewness. 

Recall that only $\Psi_{R_{1}R_{2}}(\rho_{1},\rho_{2})$  is easily computed.  The statistically relevant cumulant generating function, $\varphi_{R_{1}R_{2}}(\lambda_{1},\lambda_{2})$, is only accessible via equation~(\ref{phifromLeg})    through an inverse Legendre
transform which brings its own complications. In particular note that all values of $\lambda_{i}$ are not accessible due to the fact that 
the $\rho_{i}$ -- $\lambda_{i}$ relation cannot always be inverted. This is signaled by the fact that the determinant of the transformation vanishes,
e.g. $\Det\left[{\partial\rho_{i}\partial \rho_{j}}\Psi(\rho_{1},\rho_{2})\right]=0$. This condition is met both for finite values of 
$\rho_{i}$ and $\lambda_{i}$. 
The  corresponding contour lines of $\varphi(\lambda_{1},\lambda_{2})$ was investigated in BPC and successfully compared to simulation.

\subsubsection{Motivation}
It is beyond the scope of this letter to re-derive equations~(\ref{phifromLeg})-(\ref{FondRel}) - a somewhat detailed presentation can be found in \cite{2002A&A...382..412V} and in \cite{Bernardeau14} - 
but we can give a hint of where it comes from:
it is always possible to express any ensemble average 
in terms of the statistical properties of the initial density field  so that we can formally write
\begin{equation}
\hskip -0.1cm
\exp\left[ \varphi\right]\!\!=\!\! \int\!\!\mD\tau_{1}\mD\tau_{2}\,\mP(\tau_{1},\tau_{2})
 \exp\big(\lambda_{1}\rho_{1}(\tau_{1})+\lambda_{2}\rho_{2}(\tau_{2})\big)\,. \hskip -0.3cm
\label{phiexp2}
\end{equation}
As the present-time densities $\rho_{i}$ can arise from different initial contrasts, the above-written integration is therefore a path integral (over all the possible paths from initial conditions to present-time configuration) with measure $\mD\tau_{1}\mD\tau_{2}$ and known initial statistics $\mP(\tau_{1},\tau_{2})$. Let us assume here that the initial PDF is Gaussian so that,
\begin{equation}
 \mP(\tau_{1},\tau_{2})\dd\tau_{1}\dd\tau_{2}=
\frac{\sqrt{\det\Xi}\exp\left[-\Psi(\tau_{1},\tau_{2})\right]}{2\pi}\dd\tau_{1}\dd\tau_{2}\,,
\label{mPexp1}
\end{equation}
with $\Psi$ then a quadratic form.

In the  regime where the variance of the density field is small,
 equation~(\ref{phiexp2}) is dominated by the path corresponding to the most likely configurations.
 As the constraint is spherically symmetric, this most likely path should also respect  spherical symmetry. It is therefore bound to obey 
 the spherical collapse dynamics. Within this regime equation~(\ref{phiexp2}) becomes
  \begin{equation}
  \hskip -0.1cm
\exp\left[ \varphi\right]\!\!\simeq \!\! \int\!\! d\tau_{1}  d\tau_{2}\,\mP(\tau_{1},\tau_{2})
 \exp\big(\lambda_{1}\zeta_{{\rm SC}}(\tau_{1})\!+\!\lambda_{2}\zeta_{{\rm SC}}(\tau_{2})\big),\hskip -0.45cm
\label{mPexp3}
\end{equation}
 where the most likely path, $\rho_{i}=\zeta_{\rm SC}(\eta,\tau_{i})$ is the one-to-one spherical collapse mapping between one final density at time 
 $\eta$ and one initial density contrast as already described.  
The integration on the r.h.s. of equation~(\ref{mPexp3}) can now be carried by using 
a steepest descent method, approximating the integral as its most likely value,  where $  \lambda_{1}\rho_{1}(\tau_{1})+\lambda_{2}\rho_{2}(\tau_{2})- \Psi(\tau_{1},\tau_{2}) $ is 
stationary. It eventually leads to the fundamental relation (\ref{FondRel}) when its right hand side is computed at initial time (and the fact that 
(\ref{FondRel}) is valid for any times $\eta$ and $\eta'$ is obtained when the same reasoning is applied twice, for the two different times).

The purpose of this letter is to confront numerically computations of the two-cell PDF derived from the expression of $\varphi(\lambda_{1},\lambda_{2})$
with measurements in numerical simulations.

\begin{figure}
\center\includegraphics[width=\columnwidth]{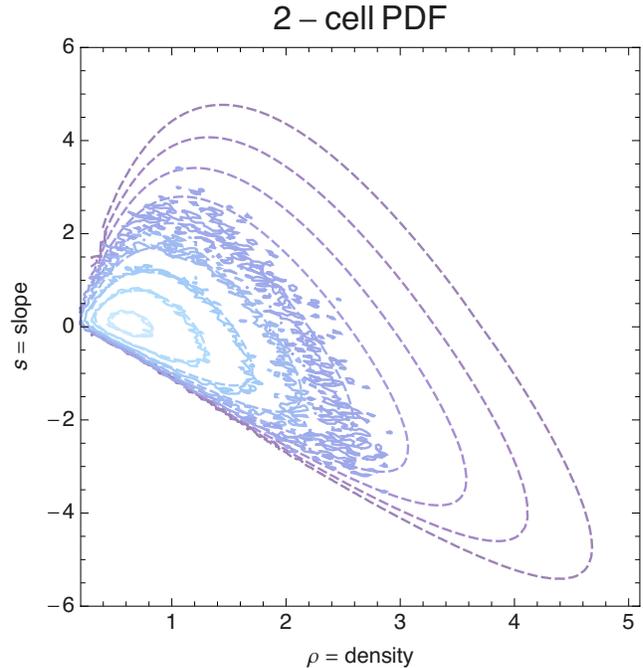}
   \caption{Joint PDF of the slope ($s$) and the density ($\rho$) as given by equation~(\ref{eq:PDF2Dinvlapmu}) for two concentric spheres of radii $R_{1}=10$ Mpc/h and $R_{2}=11$ Mpc/h at redshift $z=0.97$. Dashed contours corresponds to Log $ \mP= 0,-1/2,-1,\cdots -3$ for the theory.
   The corresponding measurements are shown as a solid line.
   \label{fig:JPDF}}
\end{figure}

\subsection{The 2-cell PDF using inverse Laplace transform}

Once the cumulant generating function is known in equation~(\ref{Mint}),
the 2-cell PDF, $\mP(\hat \rho_{1},\hat \rho_{2})$, is obtained by a 2D inverse Laplace transform of $\varphi(\lambda_{1},\lambda_{2})$
\begin{equation}
\mP\!=\!\!\int_{-\ii\infty}^{\ii\infty}\!\!\frac{\dd\lambda_{1}}{2 \pi \ii} 
\int_{-\ii\infty}^{\ii\infty}\!\!\frac{\dd\lambda_{2}}{2 \pi \ii}
\exp(-\!\!\sum_{i=1,2}\hat \rho_{i}\lambda_{i}+\varphi(\lambda_{1},\lambda_{2}))\,,
\label{eq:PDF2Dinvlap}
\end{equation}
with $\varphi$ given by equations~(\ref{phifromLeg})-(\ref{FondRel}).
From  this equation, it is straightforward to deduce the joint PDF, $\hat\mP(\hat \rho, \hat s)$, for the density, $\hrho=\hrho_1$ and 
the slope $\hat s\equiv (\hrho_2-\hrho_1) R_1/\Delta R$, $\Delta R$ being $R_{2}-R_{1}$, as 
\begin{equation}
\hat \mP=\int_{-\ii\infty}^{\ii\infty}\frac{\dd\lambda}{2 \pi \ii} 
\int_{-\ii\infty}^{\ii\infty}\frac{\dd\mu}{2 \pi \ii}
\exp(- \hat\rho \lambda - \hat s \mu+ \varphi(\lambda,\mu))\,,
\label{eq:PDF2Dinvlapmu}
\end{equation}
with $\lambda=\lambda_{1}+\lambda_{2}$, $\mu=\lambda_{2}{\Delta R}/{R_{1}}$.
  Following this definition, $ \varphi(\lambda,\mu)$ is also the Legendre transform of $\Psi(\hrho_{1},{\hat s}=(\hrho_{2}-\hrho_{1})\,R_{1}/\Delta R)$.

\begin{figure}
\includegraphics[width=\columnwidth]{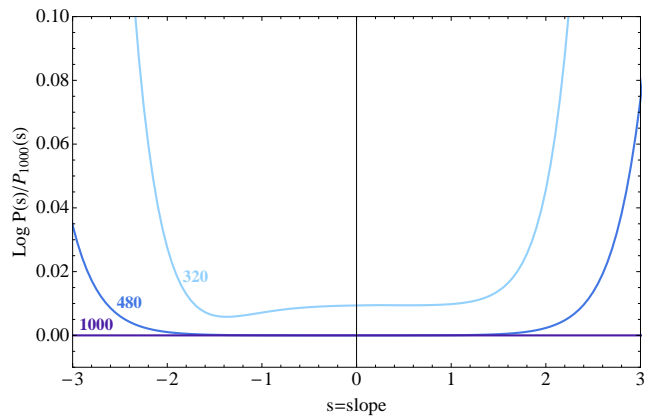}
   \caption{Dependence of the PDF of the slope on the number of points used in the numerical integration in (\ref{eq:PDF2Dinvlap}). The reference PDF is computed using $1000^{2}$ points (dark blue) and is compared to the result of the numerical integration when using $320^{2}$ (blue) and $480^{2}$ (light blue) points. 
   \label{fig:num-cvg}}
\end{figure}

In order to numerically compute equation (\ref{eq:PDF2Dinvlap}), we simply choose the imaginary path $(\lambda_{1},\lambda_{2})=\ii (n_{1}\Delta\lambda,n_{2}\Delta\lambda)$ where $n_{1}$ and $n_{2}$ are (positive or negative) integers and the step $\Delta \lambda$ has been set to 0.15. The maximum value of $\lambda_{i}$ used here is 75 resulting into a discretisation of the integrand on $1000^{2}$ points. 
Fig.~\ref{fig:JPDF} compares the result of the numerical integration of equation~(\ref{eq:PDF2Dinvlap}) to simulations.
 The corresponding dark matter simulation (carried out with {\tt Gadget2} \citep{gadget2}) is characterized by the following $\Lambda$CDM cosmology: $\Omega_{\rm m}=0.265 $, $\Omega_{\Lambda}=0.735$, $n=0.958$, $H_0=70 $ km$\cdot s^{-1} \cdot $Mpc$^{-1}$ and $\sigma _8=0.8$, 
$\Omega_{b}=0.045$
within one standard deviation of WMAP7 results \citep{wmap7}. 
The box size is 500 Mpc$/h$ sampled with  $1024^3$ particles, the softening length 24 kpc$/h$.
Initial conditions are generated using {\tt mpgrafic}  \citep{mpgrafic}.
An Octree is built  to count  efficiently all particles 
within concentric spheres of radii between $R=10$ and $ 11 {\rm Mpc}/h $.  The center of these spheres is sampled regularly on a grid of 
$ 10\, {\rm Mpc}/h $ aside, leading to 117649 estimates of the density per snapshot.
Note that the cells overlap for radii larger than $10 \, {\rm Mpc}/h $.

The convergence of our numerical scheme is investigated by varying the number of points. Fig.~\ref{fig:num-cvg} shows that the numerical integration of the slope PDF has reached 1\% precision for the displayed range of slopes. Obviously, the integration is very precise for low values of the slope and requires a largest number of points for the large-slope tails.

\section{Conditional distributions}
\begin{figure}
\includegraphics[width=\columnwidth]{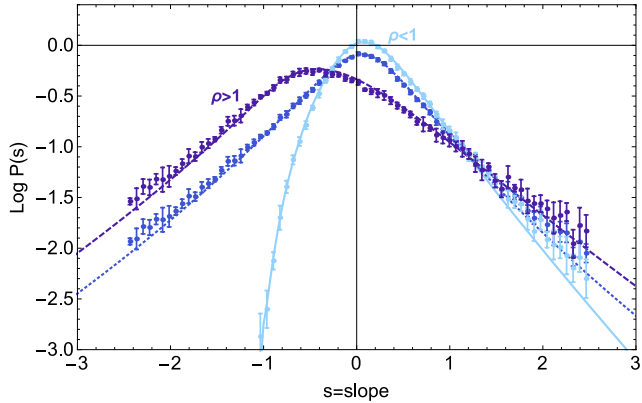}
   \caption{Density profiles in underdense (solid light blue), overdense (dashed purple) and all regions (dashed blue) for cells of radii $R_{1}=10$ Mpc/h and $R_{2}=11$ Mpc/h at redshift $z=0.97$. Predictions are successfully compared to measurements in simulations (points with error bars).
   \label{fig:profiles}}
\end{figure}

\subsection{Slope in sub regions}
Once the full 2-cell PDF is known, it is  straightforward to derive predictions for density profiles restricted to underdense 
\begin{equation}
\mP(\hat s|\hat \rho<1)=\frac{\int_{0}^{1}\dd \hat \rho\; \hat \mP(\hat\rho,\hat s)}{\int_{-\infty}^{\infty}\dd \hat s\int_{0}^{1}\dd \hat \rho\; \hat \mP(\hat\rho,\hat s)}\,,
\label{eq:slopes-over}
\end{equation}
and overdense regions
\begin{equation}
\mP(\hat s|\hat\rho>1)=\frac{\int_{1}^{\infty}\dd\hat \rho\; \hat \mP(\hat\rho,\hat s)}{\int_{-\infty}^{\infty}\dd \hat s\int_{0}^{1}\dd \hat\rho\; \hat \mP(\hat\rho,\hat s)}\,.
\label{eq:slopes-under}
\end{equation}
Fig.~\ref{fig:profiles} displays these predicted density profiles in underdense and overdense regions compared to the measurements in our simulation. 
A very good agreement is found with some slight departures in the large slope tail of the distribution.
 As expected, the underdense slope PDF peaks towards positive slope, while the overdense 
   PDF peaks towards negative slope.  The constrained negative tails are more sensitive to the
   underlying constraint, providing improved leverage for measuring the underlying cosmological parameters.   

\subsection{Density in regions of given slope}
\begin{figure}
\includegraphics[width=\columnwidth]{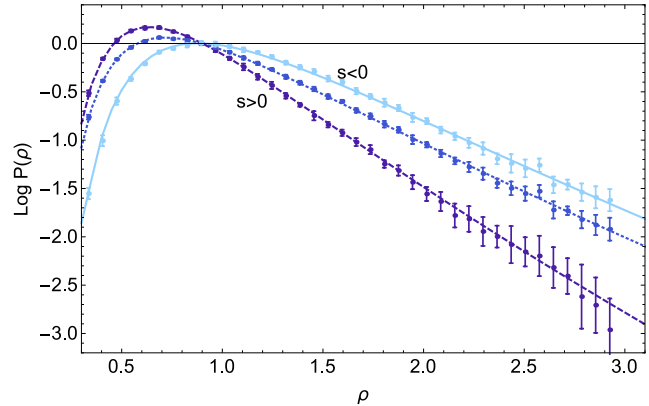}
   \caption{Density PDF in negative slope (solid light blue), positive slope (dashed purple) and all regions (dashed blue) for cells of radii $R_{1}=10$ Mpc/h and $R_{2}=11$ Mpc/h at redshift $z=0.97$. Predictions are successfully compared to measurements in simulations (points with error bars).
   \label{fig:constraineddensity}}
\end{figure}

Conversely, one can study the statistics of the density given constraints on the slope. For instance, the density PDF in regions of negative slope reads
\begin{equation}
\mP(\hat\rho|\hat s<0)=\frac{\int_{-\infty}^{0}\dd \hat s\; \hat \mP(\hat\rho,\hat s)}{\int_{0}^{\infty}\dd\hat \rho\int_{-\infty}^{0}\dd \hat s\; \hat \mP(\hat\rho,\hat s)}\,.
\label{eq:densityPDF}
\end{equation}
Fig.~\ref{fig:constraineddensity} displays the predicted density PDF in regions of positive or negative slope.
 As expected, the density is higher in regions of negative slope.  An excellent agreement with simulations is found.

\subsection{Redshift evolution}

\begin{figure}
\includegraphics[width=\columnwidth]{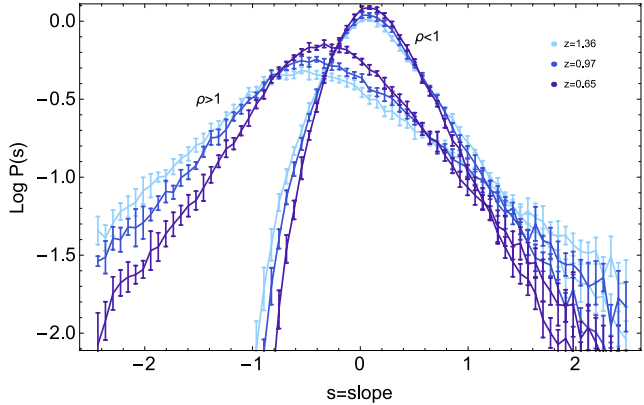}
   \caption{Same as Fig.~\ref{fig:profiles} for a range of redshifts as labeled. 
   Only the underdense  ($\rho<1$) and the overdense ($\rho>1$) PDFs are shown.
   \label{fig:profilesz}}
\end{figure}

Fig.~\ref{fig:profilesz} displays the density profiles in underdense and overdense regions as measured in the simulation
for a range of redshifts.
This figure shows that the high density subset for moderately negative slopes is particularly sensitive 
to redshift evolution, which suggests that dark energy investigations should focus on such range of slopes and regions.

\section{Fiducial dark energy experiment}
\begin{figure}
\includegraphics[width=\columnwidth]{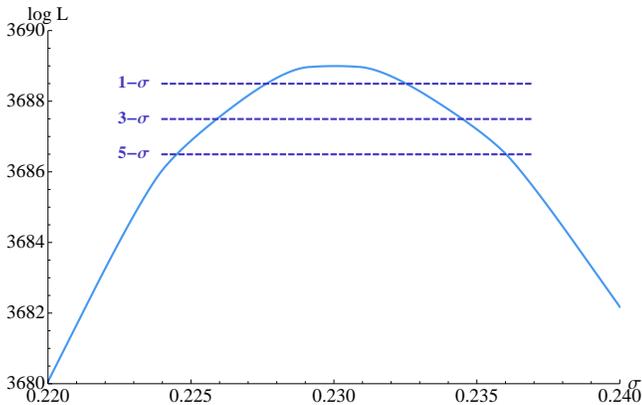}
   \caption{Log-likelihood for a fiducial experiment involving 10,000 concentric spheres 
   of 10 and 11 Mpc$/h$ measured in  our simulation. The model here only depends on the variance $\sigma$ ($\nu$ and $n$ are fixed). The contours at 1, 3 and 5 sigmas centered on the true value $0.23$ are displayed with dark blue dashed lines. The same experiment can be carried out when the three parameters vary. 
   \label{fig:likelihood}}
\end{figure}
Let us conduct the following fiducial experiment. Consider a set of 10,000 concentric spheres,
and measure for each pair the slope and the density, $\{\hat\rho_i,\hat s_i\}$.
Recall that the cosmology is encoded in the parametrization of the spherical collapse on the one hand ($\nu$),
and on the linear power-spectrum, $P^{\rm lin}_k$, (via the  covariance matrix, $\Sigma_{ij}= \int  P^{\rm lin}_k(k) W(R_i k)W(R_j k)d^3k/(2\pi)^3$ with $W(k)={3}\left(\sin(k)/k-\cos(k)\right)/{k^{2}}$) on the other hand.
For scale invariant power spectra with power index $n$, given equation~(\ref{zetarelation}), we have a three parameter $(n,\nu,\sigma)$
set of models.
 For a parametrized PDF, $P_{n,\nu,\sigma}(\hat\rho,\hat s)$ 
given by equation~(\ref{eq:PDF2Dinvlapmu}), we can compute the log likelihood of the set as 
${\cal L}(n,\nu,\sigma)=\sum_i \log P_{n,\nu,\sigma}(\hat\rho_i,\hat s_i)$. Fig.~\ref{fig:likelihood} displays the  corresponding 
likelihood contours 
at one, three and five sigmas in the simple case in which only one parameter ($\sigma$ here) varies. This experiment mimics the precision expected from a survey of useful volume of about $(350 h^{-1} \textrm{Mpc})^{3}$ which is found to be at the percent level. This work improves the findings of BPC which relies on a low-density approximation for the joint PDF.

\section{Conclusion}
Extending the analysis of BPC,
predictions for the {\sl joint} PDF of the density within  two concentric spheres was straightforwardly implemented
for a given cosmology as a function of redshift in the mildly non-linear regime. 
The agreement with measurements in simulation was shown on Figs.~\ref{fig:JPDF}, \ref{fig:profiles} and \ref{fig:constraineddensity} to be very good, 
including in the quasi-linear regime where standard perturbation theory normally fails. 
A fiducial dark energy experiment was implemented on  counts derived from $\Lambda$CDM simulations.
\\
Such statistics will prove useful in upcoming surveys as they allow us to probe differentially the slope of the density in regions of low or high density.
It can serve as a statistical indicator to test gravity and dark energy models and/or probe key cosmological parameters in carefully chosen subsets of surveys.  
The theory of count-in-cells could be applied to 2D cosmic shear maps so as to predict the statistics of projected density profiles.  Velocity profiles and combined probes involving the density and velocity fields should also be within reach of this formalism.

\vspace{0.5cm}

{\bf Acknowledgements:}  
 This work is partially supported by the grants ANR-12-BS05-0002 and  ANR-13-BS05-0005 of the French {\sl Agence Nationale de la Recherche}.
The simulations were run on 
the {\tt Horizon} cluster. 
We acknowledge support from S.~Rouberol for running the cluster for us.
\bibliographystyle{mn2e}

\bibliography{LSStructure}

\end{document}